# Particle acceleration by stimulated emission of radiation in cylinder waveguide

**TianXiu-fang(田秀芳)**[1]  **Wu Cong-feng(吴丛凤)**[1]*  **Jia Qi-ka(贾启卡)**[1]

[1](National Synchrotron Radiation Laboratory, University of Science and Technology of China, Hefei 230029, China)

**Abstract**: In the particle acceleration by stimulated emission of radiation (PASER), the efficient interaction occurs when a train of micro-bunches whose periodicity being identical to the resonance frequency of the medium. The previous theoretical calculations based on the simplified model consider only the energy exchange in the boundless condition, under the experimental condition, however, the gas active medium must be guided by the metal waveguide. In this paper, we have developed the model of energy exchange between a train of micro-bunches and gas mixture active medium in waveguide boundary for the first time based on the Theory of Electromagnetic Fields, and made detailed analysis and calculations with MathCAD. The results show that the energy density could be optimized to a certain value to get the maximum energy exchange.

**Key words:** PASER; train of micro-bunches; active gas mixture medium; cylinder waveguide;

**PACS:** 29.20.-c, 41.75.Lx

## 1 Introduction

The great demands of high-energy physics, LASER, medicine and material science spurred research into developing a new generation of compact low-cost tabletop particle accelerators with widespread applications in various fields. Particle acceleration by stimulated emission of radiation (PASER) is one of the most promising ways to achieve this goal. In PASER, energy stored in an active medium is transferred directly to the electrons traversing the medium, and therefore, accelerating the former. PASER does not need high power beam driver, and thus neither phase matching nor compensating for phase slippage is required. Moreover, electron gun is also not needed in PASER because of its capability of generating its own micro-bunched electron beam source in penning trap. Especially, with this low-cost, high compacted and simple structure PASER acceleration gradient in order of 1 GV/m could be feasibly obtained.

Since 1995, Levi Schachter and his coworkers have made a series of theory analysis and calculations about PASER process. In 2006[1], the proof of principle experiment was carried out at the Brookhaven National Laboratory Accelerator Test Facility, which features a photocathode-driven microwave linear accelerator and a high peak power $CO_2$ laser. This is the first experimental result which gave the feasibility of the PASER. Recently [2,3], Miron Voin studied the wake generated by electron bunch in $CO_2$ gas mixture active medium, and the results show that the wake accelerating gradient can be achieved of the order of GV/m.

In the PASER process, efficient interaction occurs only under the resonance condition. In 2006, Samer Banna[4] and coworkers had made theory calculation with the PASER process based on the simplified model in which they only considered the energy exchange between the train of electrons and gas mixture active medium in the boundless condition, but considering the experimental situation, the gas active medium need to be guided by the cylinder waveguide. In the following section, we developed a 2D model, deduced the formula of the energy exchange between the train of micro-bunches and the gas mixture active medium in waveguide boundary, and we made a series of analysis and calculations with MathCAD[5] on the influence of varies parameters to the gain of the micro-bunches.

[1]Supported by National Natural Science Foundation of China (10675116) and Major State Basic Research Development Programme of China (2011CB808301)
*通讯作者：吴丛凤，**E-mail**:cfwu@ustc.edu.cn
作者简介：田秀芳，女，（1983-），博士生，主要研究方向为新加速原理和新加速结构。
**E-mail**:txiufang@mail.ustc.edu.cn

## 2 Model description

We consider an electron bunch which consists of M micro-bunches[6], and assume that each micro-bunch is azimuthally symmetric and have a radius $R_b$ and length $\Delta$, carry charge $-Q$ and move at a constant velocity $v_0$. The distance between two adjacent micro-bunches is $\lambda_0$. The radius of the waveguide is $R_w$. We denote by $z_\mu$ the longitudinal coordinate of the center of the μth micro-bunch at $t=0$, as illustrated in Fig. 1. In the framework of this study, the medium is assumed to remain in the linear regime throughout the interaction region.

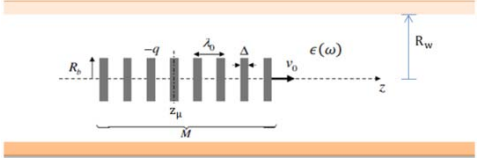

Fig.1(color online) Schematic of beam-active medium interaction in a waveguide

The moving train of micro-bunches generates a current density in the longitudinal direction z-axis that is given in the time domain by

$$J_z(r,z;t) = -qv_0 \sum_{\mu=1}^{M} \frac{1}{2\pi}\left[\frac{2}{R_b^2}h(R_b - r)\right]$$

$$\times \left\{\frac{1}{\Delta}h\left[\frac{\Delta}{2}[z - (z_\mu + v_0 t)]\right]\right\} (1)$$

For a conceptual description, we consider the wake generated by a thin charged loop ($Q_b$) of radius $R_\sigma$, located at t = 0 at $z_\sigma$ with the velocity $v_0$. The current generated by this loop is

$$J_z(r,z;t) = \frac{-Q_b v_0}{2\pi r}\delta(r - R_\sigma)\delta(z - z_\sigma - v_0 t) \ (2)$$

This current density excites the *z* component of the magnetic vector potential which, in turn, satisfies the inhomogeneous wave equation. Hence, in the frequency domain, the magnetic vector potential is given by

$$A_z(r,z;\omega)$$
$$= -\frac{\mu_0 Q_b}{(2\pi)^2}\sum_s \frac{J_0\left(p_s \frac{r}{R_w}\right)J_0\left(p_s \frac{R_\sigma}{R_w}\right)}{\frac{1}{2}R_w^2 J_1^2(p_s)}$$
$$\times \int_{-\infty}^{\infty} d\omega \frac{e^{-j\left(\frac{\omega}{v_0}\right)(z-z_\sigma)}}{-\frac{p_s^2}{R_w^2} - \frac{\omega^2}{v_0^2} + \frac{\omega^2}{c^2}\epsilon(\omega)} \quad (3)$$

So the longitudinal electric field in frequency domain is

$$E_z(r,z;\omega)$$
$$= -\frac{\mu_0 Q_b}{(2\pi)^2}\sum_s \frac{J_0\left(p_s \frac{r}{R_w}\right)J_0\left(p_s \frac{R_\sigma}{R_w}\right)}{\frac{1}{2}R_w^2 J_1^2(p_s)}$$
$$\times \int_{-\infty}^{\infty} d\omega \frac{-j\omega\left[1 - \frac{1}{\beta^2\epsilon(\omega)}\right]e^{-j\left(\frac{\omega}{v_0}\right)(z-z_\sigma)}}{-\frac{p_s^2}{R_w^2} - \frac{\omega^2}{v_0^2} + \frac{\omega^2}{c^2}\epsilon(\omega)} (4)$$

Based on the Green theory, the electric field generated by the train of the micro-bunches is expressed by

$$E_z(r,z;\omega)$$
$$= -\frac{\mu_0 q}{(2\pi)^2}\sum_{\mu=1}^{M}\sum_s \frac{J_c^2\left(p_s \frac{R_b}{R_w}\right)}{\frac{1}{2}R_w^2 J_1^2(p_s)}\text{sinc}\left(\frac{\omega\Delta}{2v_0}\right)$$
$$\times \int_{-\infty}^{\infty} d\omega \frac{\frac{c^2}{j\omega\epsilon(\omega)}\left[\frac{\omega^2}{c^2}\epsilon(\omega) - \frac{\omega^2}{v^2}\right]}{-\frac{p_s^2}{R_w^2} - \frac{\omega^2}{v^2} + \frac{\omega^2}{c^2}\epsilon(\omega)}e^{-j\left(\frac{\omega}{v_0}\right)(z-z_\mu)} (5)$$

Using the Fourier transform, we can get the electric field in time domain

$$E_z(r,z;t)$$
$$= -\frac{\mu_0 q}{(2\pi)^2}\sum_{\mu=1}^{M}\sum_s \frac{J_c^2\left(p_s \frac{R_b}{R_w}\right)}{\frac{1}{2}R_w^2 J_1^2(p_s)}\text{sinc}\left(\frac{\omega\Delta}{2v_0}\right)$$
$$\times \int_{-\infty}^{\infty} d\omega \frac{\frac{c^2}{j\omega\epsilon(\omega)}\left[\frac{\omega^2}{c^2}\epsilon(\omega) - \frac{\omega^2}{v^2}\right]}{-\frac{p_s^2}{R_w^2} - \frac{\omega^2}{v^2} + \frac{\omega^2}{c^2}\epsilon(\omega)}e^{j\omega\left(t-\frac{z-z_\mu}{v_0}\right)} （6）$$

Using the term for the longitudinal electric field as determined in Eq.(6), and the expression for the current density Eq. (1), and integrating the power density over the beam's volume, we get the total power exchange in the interaction process.

$$P(t) = -2\pi \int_0^\infty rdr \int_{-\infty}^\infty J_z(r,z;t)E_z(r,z;t)dz$$

$$= -2\pi \int_0^\infty rdr \int_{-\infty}^\infty dz qv_0 \sum_{\mu'=1}^M \frac{1}{2\pi}\left[\frac{2}{R_b^2}h(R_b-r)\right]$$

$$\times \left\{\frac{1}{\Delta}h\left[\frac{\Delta}{2}[z-(z_{\mu'}+v_0t)]\right]\right\}$$

$$\times \frac{\mu_0 q}{(2\pi)^2}\sum_{\mu=1}^M \sum_s \frac{J_c^2\left(p_s\frac{R_b}{R_w}\right)}{\frac{1}{2}R_w^2 J_1^2(p_s)} sinc\left(\frac{\omega\Delta}{2v_0}\right)$$

$$\times \int_{-\infty}^\infty d\omega \frac{\frac{c^2}{j\omega\epsilon(\omega)}\left[\frac{\omega^2}{c^2}\epsilon(\omega)-\frac{\omega^2}{v^2}\right]}{-\frac{p_s^2}{R_w^2}-\frac{\omega^2}{v_0^2}+\frac{\omega^2}{c^2}\epsilon(\omega)}e^{j\omega\left(t-\frac{z-z_\mu}{v_0}\right)} \quad (7)$$

Integrating on the horizontal and vertical respectively, we can obtain the simplified expression as follows

$$P(t)$$

$$= -2\pi qv_0 \frac{1}{2\pi}\frac{\mu_0 q}{(2\pi)^2}\sum_{\mu'=1}^M \sum_{\mu=1}^M \sum_s \frac{J_c^2\left(p_s\frac{R_b}{R_w}\right)}{\frac{1}{2}R_w^2 J_1^2(p_s)}$$

$$\times sinc\left(\frac{\omega\Delta}{2v_0}\right)\int_{-\infty}^\infty d\omega \frac{\frac{c^2}{j\omega\epsilon(\omega)}\left[\frac{\omega^2}{c^2}\epsilon(\omega)-\frac{\omega^2}{v^2}\right]}{-\frac{p_s^2}{R_w^2}-\frac{\omega^2}{v_0^2}+\frac{\omega^2}{c^2}\epsilon(\omega)}$$

$$\times e^{-j\omega\frac{z_{\mu'}-z_\mu}{v_0}}sinc\left(\frac{\omega\Delta}{2v_0}\right) \quad (8)$$

Wherein $z_\mu = z_1 + (\mu-1)\lambda_0$, $\mu = 1,2,...,M$, $\eta_0 = \sqrt{\mu_0/\epsilon_0}$ is the free space impedence. The double averaging over the longitudinal initial location of the micro-bunches may be simplified to read

$$\sum_{\mu'=1}^M \sum_{\mu=1}^M e^{-j\left(\frac{\omega}{v_0}\right)(z_{\mu'}-z_\mu)} = M^2 \frac{sinc^2\left(\frac{\omega\lambda_0 M}{2v_0}\right)}{sinc^2\left(\frac{\omega\lambda_0}{2v_0}\right)} \quad (9)$$

So the total power exchange is

$$P(t)$$

$$= -\frac{Q^2 v_0 \mu_0}{2\pi^2 R_w^2}\sum_s \frac{J_c^2\left(p_s\frac{R_b}{R_w}\right)}{J_1^2(p_s)}$$

$$\times \int_{-\infty}^\infty d\omega \frac{\frac{c^2}{j\omega\epsilon(\omega)}\left[\frac{\omega^2}{c^2}\epsilon(\omega)-\frac{\omega^2}{v^2}\right]}{-\frac{p_s^2}{R_w^2}-\frac{\omega^2}{v_0^2}+\frac{\omega^2}{c^2}\epsilon(\omega)}$$

$$\times M^2 \frac{sinc^2\left(\frac{\omega\lambda_0 M}{2v_0}\right)}{sinc^2\left(\frac{\omega\lambda_0}{2v_0}\right)}sinc^2\left(\frac{\omega\Delta}{2v_0}\right) \quad (10)$$

So the energy exchange during the passage d is

W

$$= \frac{Q^2 d\eta_0^2}{2\pi^2 R_w^2}\sum_s \frac{J_c^2\left(p_s\frac{R_b}{R_w}\right)}{J_1^2(p_s)}$$

$$\times \int_{-\infty}^\infty d\omega \frac{j\omega\left[\epsilon_r(\omega)-\frac{1}{\beta^2}\right]}{\epsilon_r(\omega)\left[-\frac{p_s^2}{R_w^2}-\frac{\omega^2}{v_0^2}+\frac{\omega^2}{c^2}\epsilon_r(\omega)\right]}$$

$$\times M^2 \frac{sinc^2\left(\frac{\omega\lambda_0 M}{2v_0}\right)}{sinc^2\left(\frac{\omega\lambda_0}{2v_0}\right)}sinc^2\left(\frac{\omega\Delta}{2v_0}\right) \quad (11)$$

Considering the gas mixture active medium which is represented by a dielectric function

$$\epsilon_r(\omega>0) \equiv 1 + \frac{\omega_p^2}{\omega_0^2-\omega^2+2j\omega/T_2} \quad (12)$$

It is convenient to introduce a set of normalized quantities as follows:

$$\bar{R}_b = \frac{R_b}{\lambda_0}, \bar{\Delta} = \frac{\Delta}{\lambda_0}, \Omega = \omega\frac{\lambda_0}{c}, \varphi = \frac{\Omega}{2\beta}, Q_{total} = MQ$$

$$F_\parallel(\varphi,\bar{\Delta},M) = sinc^2(\varphi\bar{\Delta})\frac{sinc^2(\varphi M)}{sinc^2(\varphi)}$$

$$F_\perp(\Omega) = \frac{1}{\left[-\frac{p_s^2}{R_w^2}-\frac{\Omega^2}{\lambda_0^2\beta^2}+\frac{\omega^2}{c^2}\epsilon_r(\Omega)\right]}$$

Then the normalized dielectric function is:

$$\epsilon_r(\Omega>0) = 1 + \frac{\Omega_p^2}{(2\pi)^2-\Omega^2+2j\alpha\Omega}$$

In which

$$\alpha = \lambda_0/cT_2, \quad \Omega_p^2 = \omega_p^2\lambda_0^2/c^2$$

And evaluate the integral only at the poles of the dielectric function

$$\epsilon_r(\Omega) = 0 \quad (13)$$

We obtain

$$\Omega = \Omega_\pm = j\alpha \pm \sqrt{(2\pi)^2 + \Omega_p^2 - \alpha^2}$$

$$= j\alpha \pm \Omega_R \quad (14)$$

Wherein $\Omega_R = \sqrt{(2\pi)^2 + \Omega_p^2 - \alpha^2}$.

Using Cauchy residue theorem, the energy exchange reads

$$W = -\frac{Q_{total}{}^2 d\eta_0{}^2 c}{\pi R_w^2 \lambda_0 \beta^2} \frac{\Omega_p^2}{\Omega_R} \sum_s \frac{J_c^2\left(p_s \frac{R_b}{R_w}\right)}{J_1^2(p_s)}$$

$$\times F_\parallel\left(\frac{\Omega_R}{2\beta}, \bar{\Delta}, M\right) \text{Re}[\Omega_+ F_\perp(\Omega_+)] \quad (15)$$

So the energy obtained by the electron micro-bunches $\Delta E_k = -W$ is

$$\Delta E_k = \frac{Q_{total}{}^2 d\eta_0{}^2 c}{\pi R_w^2 \lambda_0 \beta^2} \frac{\Omega_p^2}{\Omega_R} \sum_s \frac{J_c^2\left(p_s \frac{R_b}{R_w}\right)}{J_1^2(p_s)}$$

$$\times F_\parallel\left(\frac{\Omega_R}{2\beta}, \bar{\Delta}, M\right) \text{Re}[\Omega_+ F_\perp(\Omega_+)] \quad (16)$$

And the relative energy change is expressed by

$$\Delta \bar{E}_k = \frac{\Delta E_k}{N_{el} mc^2 (\gamma - 1)} \quad (17)$$

The expression in equation (16) shows that the energy exchange between the train of electron micro-bunches and the gas active medium is related to the number of modes and is linearly proportion to interaction length d.

## 3 Influencing of the parameters to the energy exchange

Similar to the boundless condition, we consider the influence of various parameters to the energy exchange in the waveguide.

Fig.2 illustrates the relative change in the kinetic energy of the macro-bunch versus the energy density stored in the excited CO2 gas mixture at the resonance frequency when considering 500 modes. $R_w = 120\lambda_0$, $R_b = 10\lambda_0$, $\Delta = 0.1\lambda_0$, M=150, In fig.2a is the result for the waveguide boundary condition, while fig.2b for the boundless condition.

From the simulation result in fig.2, we can see that the kinetic energy of the train of micro-bunches in the boundary condition is smaller than that in the boundless condition[7], and energy gain also oscillates as a function of $w_{act}$, and when the energy density stored in the excited $CO_2$ gas mixture medium is $w_{act} = 1550 J/m^3$, the largest energy gain can be obtained. So the optimum energy density is $w_{act} = 1550 J/m^3$.

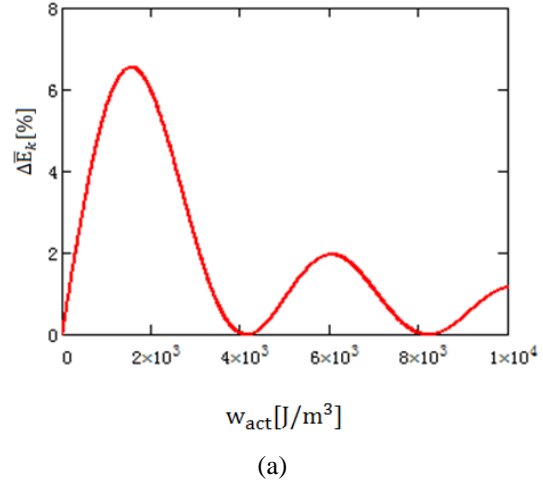

(a)

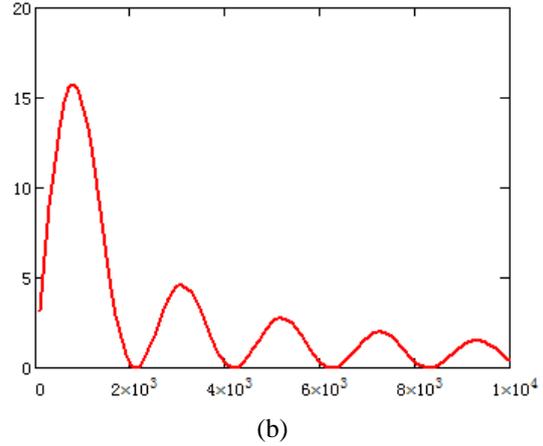

(b)

**Fig.2** (color online) the relative change in the kinetic energy of the macro-bunch versus the energy density stored in the excited $CO_2$ gas mixture

(a) Waveguide boundary condition

(b) Boundless condition

Fig.3 shows the influence of the radius of the waveguide to the energy exchange.

In fig.3, the result illustrates that with the increase of the radius of the waveguide, the effect of the radius to the energy exchange is small, when the radius is large enough, the energy exchange is no longer exchange to the radius which is corresponding to the boundless condition.

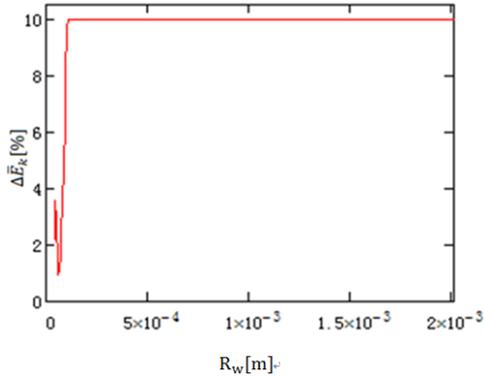

**Fig.3**(color online)  the relative change in the kinetic energy of the macro-bunch versus the radius of the waveguide

In fig.4, we show the influence of the average initial kinetic energy in the macro-bunch to energy exchange.

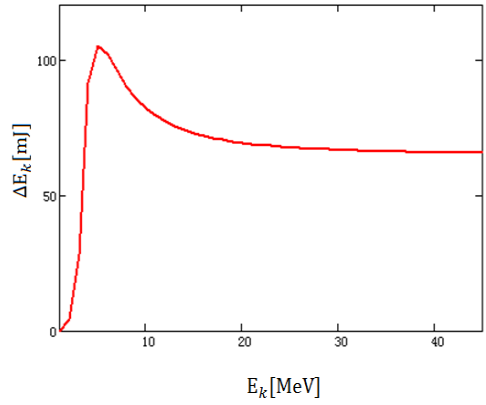

**Fig.4** (color online) The macro-bunch kinetic energy increase versus the average kinetic energy of the electrons.

For relativistic electrons, It illustrates that the kinetic energy increase of the electron beam is $\gamma$ independent, as any viable acceleration structure must exhibit. In the former calculation, we let $E_k = 45\text{MeV}$ which satisfy the conditions theory of relativity, so the kinetic energy increase doesn't change with the initial kinetic energy.

In Fig.5 we show the kinetic energy gain versus the length of a single micro-bunch, the energy gain vanishes as the length of the micro-bunch approaches the modulation wavelength and it reaches a maximum for the shortest bunch. In practice, the length of the micro-bunch is limited primarily by the modulation process and by the space-charge effects within the micro-bunch.

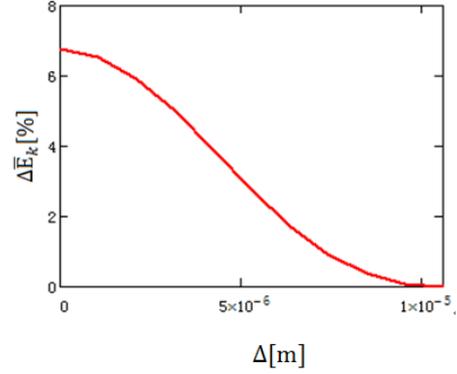

**Fig.5**(color online) The macro-bunch kinetic energy increase versus the length of the micro-bunch

The kinetic energy gain for a given amount of charge in the micro-bunch versus the number of micro-bunches was examined, and the result is presented in Fig. 6.

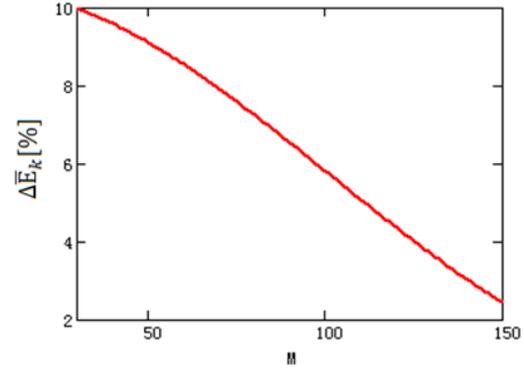

**Fig.6**(color online) The macro-bunch kinetic energy increase versus the number of micro-bunches

When the charge in the macro-bunch is fixed, with the increase of the number of micro-bunches, the total charge in the macro-bunch decreases, and so is the decrease of the energy gain.

## 4 Conclusion

The theory and simulation results about energy exchange between the train of micro-bunches and the gas mixture active medium in a waveguide above show that the train of electrons moving in an active medium in waveguide condition can get energy from the active medium, and Because the PASER process doesn't need phase matching or compensate of phase slipping, so we can extend the interaction length at will to get enough energy we need.